
%
%
%
\def\bra#1{{\langle#1\vert}}
\def\ket#1{{\vert#1\rangle}}

\def\sst#1{{\scriptscriptstyle #1}}

\def\rbra#1{{\langle #1 \vert\!\vert}}
\def\rket#1{{\vert\!\vert #1\rangle}}

\def\alr{{A_\sst{LR}}}

\def\evec{{\vec e}}

\def\mn{{m_\sst{N}}}
\def\mns{{m^2_\sst{N}}}

\def\sbar{{\bar s}}

\def\sstw{{\sin^2\theta_\sst{W}}}

\def\tauv{{\vec\tau}}
\def\sigv{{\vec\sigma}}

\def\sqr#1#2{{\vcenter{\vbox{\hrule height.#2pt
                \hbox{\vrule width.#2pt height#1pt \kern#1pt
                        \vrule width.#2pt}
                \hrule height.#2pt}}}}
\def\square{{\mathchoice\sqr74\sqr74\sqr{6.3}3\sqr{3.5}3}}

\def\sst#1{{\scriptscriptstyle #1}}

\def\mpi{{m_\pi}}
\def\mpis{{m^2_\pi}}

\def\mn{{m_\sst{N}}}
\def\mns{{m^2_\sst{N}}}

\def\mro{{m_\rho}}
\def\mros{{m^2_\rho}}

\def\gpnn{{g_{\sst{NN}\pi}}}
\def\grnn{{g_{\sst{NN}\rho}}}

\def\sst#1{{\scriptscriptstyle #1}}

\def\That{{\hat T}}

\def\Mhat{{\hat M}}

\def\mn{{m_\sst{N}}}
\def\mns{{m_\sst{N}^2}}

\def\mpi{{m_\pi}}
\def\mpis{{m^2_\pi}}

\def\sigv{{\vec\sigma}}
\def\tauv{{\vec\tau}}

\def\gpnn{{g_{\pi\sst{NN}}}}

\def\rbra#1{{\langle#1\parallel}}
\def\rket#1{{\parallel#1\rangle}}

\def\xivz{{\xi_\sst{V}^{(0)}}}

\def\xivtez{{\xi_\sst{V}^{T=0}}}

\def\GES{{G_\sst{E}^{(s)}}}
\def\GMS{{G_\sst{M}^{(s)}}}

\def\mustr{{\mu_s}}

\def\rhostr{{\rho_s}}

\def\GETEZ{{G_\sst{E}^{\sst{T}=0}}}

\def\bra#1{{\langle#1\vert}}
\def\ket#1{{\vert#1\rangle}}

\def\sst#1{{\scriptscriptstyle #1}}

\def\rbra#1{{\langle #1 \vert\!\vert}}
\def\rket#1{{\vert\!\vert #1\rangle}}

\def\alr{{A_\sst{LR}}}

\def\evec{{\vec e}}

\def\mn{{m_\sst{N}}}
\def\mns{{m^2_\sst{N}}}

\def\sbar{{\bar s}}

\def\sstw{{\sin^2\theta_\sst{W}}}

\def\tauv{{\vec\tau}}
\def\sigv{{\vec\sigma}}

\def\sqr#1#2{{\vcenter{\vbox{\hrule height.#2pt
                \hbox{\vrule width.#2pt height#1pt \kern#1pt
                        \vrule width.#2pt}
                \hrule height.#2pt}}}}
\def\square{{\mathchoice\sqr74\sqr74\sqr{6.3}3\sqr{3.5}3}}

\def\sst#1{{\scriptscriptstyle #1}}

\def\mpi{{m_\pi}}
\def\mpis{{m^2_\pi}}

\def\mn{{m_\sst{N}}}
\def\mns{{m^2_\sst{N}}}

\def\mro{{m_\rho}}
\def\mros{{m^2_\rho}}

\def\gpnn{{g_{\sst{NN}\pi}}}
\def\grnn{{g_{\sst{NN}\rho}}}

\def\sst#1{{\scriptscriptstyle #1}}

\def\That{{\hat T}}

\def\Mhat{{\hat M}}

\def\mn{{m_\sst{N}}}
\def\mns{{m_\sst{N}^2}}

\def\mpi{{m_\pi}}
\def\mpis{{m^2_\pi}}

\def\sigv{{\vec\sigma}}
\def\tauv{{\vec\tau}}

\def\gpnn{{g_{\pi\sst{NN}}}}

\def\rbra#1{{\langle#1\parallel}}
\def\rket#1{{\parallel#1\rangle}}

\def\xivz{{\xi_\sst{V}^{(0)}}}

\def\xivtez{{\xi_\sst{V}^{T=0}}}

\def\GES{{G_\sst{E}^{(s)}}}
\def\GMS{{G_\sst{M}^{(s)}}}

\def\mustr{{\mu_s}}

\def\rhostr{{\rho_s}}

\def\GETEZ{{G_\sst{E}^{\sst{T}=0}}}

\def\PRC#1{{\it Phys. Rev.} {\bf C#1} }
\def\PRD#1{{\it Phys. Rev.} {\bf D#1} }

\def\NPA#1{{\it Nucl. Phys.} {\bf A#1} }
\def\NPB#1{{\it Nucl. Phys.} {\bf B#1} }
\def\AoP#1{{\it Ann. of Phys.} {\bf #1} }

\def\PLB#1{{\it Phys. Lett.} {\bf B#1} }

\def\ZPC#1{{\it Z. f\"ur Phys.} {\bf C#1} }
\def\etal{{\it et al.}}

\def\mv{{m_\sst{V}}}
\def\gropi{{g_{\rho\pi}^{(s)}}}

\hfuzz=50pt

\vsize=7.5in
\hsize=5.6in
\magnification=1200
\tolerance 10000
\input boardmacs

\baselineskip 12pt plus 1pt minus 1pt
\pageno=0
\centerline{\bf MESON-EXCHANGE CURRENTS AND THE}
\smallskip
\centerline{{\bf STRANGENESS RADIUS OF $^4$He}\footnote{*}{This
work is supported in part by funds
provided by the U. S. Department of Energy (D.O.E.) under contracts
\#DE-AC05-84ER40150 and DE-AC02-76ER03069.}}
\vskip 24pt
\centerline{M. J. Musolf\footnote{**}{N.S.F. Young Investigator}}
\vskip 12pt
\centerline{\it Department of Physics}
\centerline{\it Old Dominion University}
\centerline{\it Norfolk, VA\ \ 23529\ \ \  U.S.A.}
\centerline{\it and}
\centerline{\it CEBAF Theory Group, MS 12H}
\centerline{\it Newport News, VA\ \ 23606\ \ \ U.S.A.}
\vskip 12pt
\centerline{\it and}
\vskip 12pt
\centerline{T.W. Donnelly}
\vskip 12pt
\centerline{\it Center for Theoretical Physics}
\centerline{\it Laboratory for Nuclear Science}
\centerline{\it and Department of Physics}
\centerline{\it Massachusetts Institute of Technology}
\centerline{\it Cambridge, Massachusetts\ \ 02139\ \ \ U.S.A.}
\vfill
\noindent CEBAF\#TH-93-16\ and\ MIT CTP\#2234\hfill September, 1993
\eject
\baselineskip 16pt plus 2pt minus 2pt
\centerline{\bf ABSTRACT}
Meson-exchange current contributions to the strangeness radius of $^4$He
are computed in the one-boson exchange approximation. It is found that
these contributions introduce a $\lapp$10\% correction to the one-body
contribution. They should not, therefore, hamper the extraction of the
nucleon strangeness radius from the parity-violating electron-$^4$He
asymmetry.
\vfill
\eject
        There has been considerable interest recently in the use of
intermediate-energy semileptonic scattering to study the strange-quark
content of the nucleon [1-10]. The elastic neutrino-nucleon and
anti-neutrino nucleon cross sections are particularly sensitive to the
nucleon's strange-quark axial vector form factor [1,9,10]. The parity-violating
(PV) elastic electron-proton and electron-nucleus asymmetries, on the other
hand, can be sensitive to the nucleon's strange-quark vector current matrix
elements [2-9]. The latter are parameterized by two
form factors: the strangeness electric ($\GES$) form factor, which
must vanish at $Q^2=0$ since the nucleon has no net strangeness,
and the strangeness magnetic ($\GMS$) form factor. Accordingly the
leading $Q^2$-dependence of the former may be characterized
by a mean-square \lq\lq strangeness radius" [11], which one may
define as a dimensionless parameter $\rhostr$ [7,9]:
$$
\rhostr={d\GES\over d\tau}\Bigr\vert_{\tau=0}\ \ \ ,\eqno(1)
$$
where $\tau = |Q^2|/4\mns$ (here, $Q^2=q^2-\omega^2\leq 0$ is the square of
the four-momentum transfer with $q=|{\vec q}|$ the three-momentum transfer
and $\omega$ the energy transfer). Similarly, one defines a strange magnetic
moment as $\mustr=\GMS(0)$.

        Ideally, one would attempt to determine $\rhostr$
and $\mustr$ with a series of PV electron scattering experiments using a proton
target. As discussed elsewhere, however, such a strategy would not necessarily
permit a separation of these two parameters at the level of precision needed
to distinguish among theoretical models [7-9, 11-16].
In principle, augmenting a PV
$\evec p$ scattering program with measurements of the PV asymmetry for
scattering from a nucleus could permit a more precise determination of
$\rhostr$ and $\mustr$ [7-9]. To that end, the $(J^\pi,T)=(0^+,0)$ nuclei
represent an attractive case. The PV asymmetry for elastic scattering from
such targets depends on
the ratio of isoscalar weak neutral current (NC) and isoscalar electromagnetic
(EM) Coulomb matrix elements. To a large extent,
the dependence of these matrix elements on
details of the nuclear wave function cancels out from their ratio,
making the asymmetry
primarily sensitive to electroweak couplings and single-nucleon electric
form factors. For purposes of extracting information on the nucleon's
strangeness radius, one would like to estimate the scale of nuclear corrections
to the $(0^+,0)$ asymmetry. In this note, we investigate one class of nuclear
corrections --- meson-exchange currents (MEC) --- to the strangeness radius
of $^4$He,
a nucleus which will be used as a target in up-coming CEBAF experiments
[17-19].
We find that, in the one-boson-exchange approximation, these corrections are
sufficiently small so as not to introduce serious uncertainty in a
determination of $\rhostr$ from the $^4$He asymmetry. MEC
contributions to the non-leading $Q^2$-behavior of the $^4$He strangeness
form factor will be discussed in a forthcoming publication [20].

        The PV asymmetry for scattering from a $(0^+,0)$ nucleus is given
by [7-9]
$$
\alr = -{G_\mu\vert Q^2\vert\over 4\sqrt{2}\pi\alpha}\left\{\sqrt{3}
\xivtez+ \xivz{\rbra{\hbox{g.s.}}\Mhat_0(s)\rket{\hbox{g.s.}}\over
\rbra{\hbox{g.s.}}\Mhat_0(T=0)\rket{\hbox{g.s.}}}\right\}\ \ \ ,\eqno(2)
$$
where $G_\mu$ is the Fermi constant measured in muon decay, $\alpha$ the
EM fine structure constant, and $\xivtez$ and $\xivz$ isoscalar and
SU(3)-singlet
vector current couplings of the $Z^0$ to the nucleon, respectively [7-9].
The operators $\Mhat_0(s)$ and $\Mhat_0(T=0)$ are
the Coulomb multipole projections of the strange-quark
vector current ($\sbar\gamma_\mu s$) and isoscalar EM current, respectively,
and
$\rbra{\hbox{\rm g.s.}}\ \ \rket{\hbox{\rm g.s.}}$ denotes a ground state
reduced
matrix element. At tree level in the Standard Model, the electroweak couplings
are $\sqrt{3}\xivtez=-4\sstw$ and $\xivz=-1$.
In the one-body approximation, the nuclear operators contained in
the multipole projections are identical, apart from the single-nucleon form
factors, so that the ratio of their matrix elements becomes
$$
{\rbra{\hbox{g.s.}}\Mhat_0(s)\rket{\hbox{g.s.}}\over
\rbra{\hbox{g.s.}}\Mhat_0(T=0)\rket{\hbox{g.s.}}}\biggr\vert_{\rm 1-body}
=\quad {\GES(Q^2)\over\GETEZ(Q^2)}\ \ \ ,\eqno(3)
$$
independent of the details of the nuclear wave functions. At low momentum
transfers, one has $\GETEZ=1/2 +{\cal O}(\tau)$ and $\GES=\rhostr\tau
+{\cal O}(\tau^2)$, so that the scale of the second term in Eq.~(2) is
essentially determined by the nucleon's strangeness radius.

        Meson-exchange current corrections to this result are generated
by processes illustrated in the diagrams of Figs. 1 and 2. Inclusion of the
currents in Fig. 1, where the exchanged meson is typically one of the
lowest-lying pseudoscalar or vector mesons, is required in order to maintain
consistency between the nucleon-nucleon interaction and conservation of EM
charge and baryon number. For isoscalar vector currents, such as
$J_\mu^\sst{EM}(T=0)$ and $\sbar\gamma_\mu s$, only the \lq\lq pair current"
($N\bar N$) processes of Fig. 1a,b contribute. The amplitude associated with
the \lq\lq meson-in-flight" process of Fig. 1c vanishes,
since by G-parity one
has that $\bra{M}V_\mu(T=0)\ket{M}=0$ for any meson $M$ and
isoscalar neutral current $V_\mu(T=0)$. When the exchanged meson
is a pion and pseudovector coupling is used at the
$\pi NN$ vertices, the corresponding two-body
isoscalar EM and strangeness Coulomb operators go like $q^2 F_1(T=0)$
and $q^2F_1^{(s)}\sim q^4$, respectively, where $F_1$ is the Dirac form
factor of the nucleon.
Hence, the longest-range MEC's -- those resulting from single
$\pi$-exchange -- will not
contribute to the nuclear strangeness radius. In the case of the $\rho$- and
$\omega$-meson
exchanges, the corresponding two-body isoscalar Coulomb operators are
also of ${\cal O}(q^2)$ times a linear combination of the electric and magnetic
nucleon form factors. Thus, when analyzing the leading $|Q^2|$-dependence of
the ratio in Eq.~(3), we may neglect
these contributions to the EM matrix element appearing in the denominator,
since they will only contribute to that ratio in ${\cal O}(q^4)$.
In contrast, the
nuclear strangeness radius (the numerator of Eq.~(3)) receives a contribution
from these currents, since $\GMS$ does not necessarily vanish at the photon
point. The resultant two-body operator has the form
$$
\eqalign{
\Mhat_0^{V}(s)\Bigr\vert_{q\to 0}& = \tau\sum_{i<j}\left[
{g_\sst{V}^2 m_\sst{V}\over 24\pi^{3/2}\mn}
\right]\That_\sst{V}(i,j)(1+\kappa_\sst{V})\cr
&\cr
&\qquad\times\left[{e^{-\mv r}\over \mv r}\right](1+\mv r)\left[1+
{2\over 3}\sigv_i\cdot\sigv_j +\cdots\right]\ \ \ ,\cr}\eqno(4)
$$
where $V=\rho$ or $\omega$; $g_\sst{V}\equiv g_\sst{VNN}$ gives the
vector meson-nucleon coupling;
$r$ is the relative separation of the two nucleons; the sum is performed
over all distinct pairs of nucleons labelled $(i,j)$; and
$$
\That_\sst{V}(i,j)=\cases{\tauv_i\cdot\tauv_j,& $V=\rho$\cr
                          1,& $V=\omega$\ \ \ \ .\cr}\eqno(5)
$$
In writing Eq.~(4) we have neglected terms ($+\cdots$) which give
non-vanishing matrix elements only if the nucleon pair carries orbital angular
momentum $L>0$ either internally or with respect to the nuclear center of
mass. For the case of $^4$He, these terms only contribute to the Coulomb
matrix elements via configuration mixing ({\it e.g.}, D-state admixtures
into the ground state), and therefore may be neglected for the present
purposes of setting a scale.

        Isobar currents of the type shown in Fig. 2a contribute
to the nuclear strangeness radius only when the intermediate nucleon
resonance carries the same isospin as the nucleon, since
the strange-quark vector current is a  (strong) isospin singlet operator. The
lightest $T=1/2$ resonance is the $N(1440)$. Although we have not computed
the contribution of the corresponding isobar current explicitly, we neglect
it on the assumption that it is suppressed by the large mass splitting between
this resonance and the nucleon.

        A third class of MEC contributions -- the so-called \lq\lq transition
currents" illustrated in Fig. 2b -- involves the matrix element of $\sbar
\gamma_\mu s$ between two different mesons, $M$ and $M'$. The lightest such
pair is the $\rho$ and $\pi$, whose strange vector current matrix element
may be written as
$$
\bra{\rho^a(k_1,\varepsilon)}\sbar\gamma_\mu s\ket{\pi^b(k_2)}
={\gropi(Q^2)\over\mro}\epsilon_{\mu\nu\alpha\beta}k_1^\nu k_2^\alpha
\varepsilon^{\beta\ast}\delta_{ab}\ \ \ ,\eqno(6)
$$
where $\varepsilon^\beta$ is the $\rho$-meson polarization and $(a,b)$
are isospin indices. Note that there is no analogous $\omega-\pi$ matrix
element since the strange vector current is isoscalar. In contrast to
the situation with diagonal $\sbar\gamma_\mu s$ currents, no conservation
principle requires the form factor $\gropi(Q^2)$ to vanish at the photon
point. Consequently, the {\it nuclear} Coulomb matrix element of the
transition charge operator derived from diagram 2b can, in principle,
contribute to the nuclear strangeness radius. Unlike the $\rho$-meson
pair-current
contribution embodied in Eq.~(4), the $\rho - \pi$ transition contribution
bears no connection with the nuclear potential used in computing nuclear
wave functions. It thus generates an unambiguous many-body correction to
the one-body form of the PV asymmetry. The corresponding two-body Coulomb
operator is, to leading order in $q^2$,
$$
\eqalign{\Mhat_0^{\rho\pi}(s)\Bigr\vert_{q\to 0}&=-\tau
\sum_{i<j}\left({2\over 9}\right) \left[
{\gpnn g_{\rho\sst{NN}} \gropi\over\pi^{3/2}}\right]\tauv_i\cdot\tauv_j\cr
&\cr
&\quad\times (1+\kappa_\rho){1\over\mro
r}\left({1\over\mros-\mpis}\right)\left[
\mpis e^{-\mpi r}-\mros e^{-\mro r}\right]\sigv_i\cdot\sigv_j +\cdots
\ \ \ ,\cr}\eqno(7)
$$
where terms with vanishing matrix elements in a ground state of pure S-waves
have been omitted. In obtaining the two-body operators in Eqs.~(4) and (7),
we did not include form factors and the hadronic vertices. For purposes
of obtaining an upper bound on the scale of MEC corrections, however,
the use of point meson-nucleon couplings should be sufficient.
For comparison, the one-body strangeness Coulomb
operator has the low-$q^2$ form
$$
\Mhat_0^{(1)}(s)\Bigr\vert_{q\to 0}={\tau\over 2\sqrt{\pi}}\rhostr\sum_i^A 1
\ \ \ .\eqno(8)
$$

        To set the scale of the different one- and two-body contributions to
the strangeness radius of $^4$He, we computed matrix elements of the
operators in Eqs.~(4,7,8) using a simple  $^4$He ground state
wave function consisting of
a Slater determinant of harmonic oscillator single-particle S-state
wave functions. Since the one-body operator in Eq.~(8) does not probe
the nuclear wave function but simply counts the number of nucleons, its
matrix element is not dependent on our choice of wave function. The two-body
matrix elements, in contrast, are more sensitive to the choice of wave
function.
A particularly important consideration in this respect is the role played
by short-range nucleon-nucleon anti-correlations, since all of the MEC's
treated
here involve at least one heavy vector meson propagator. The pieces of
the resultant two-body operators associated with the heavy meson will have
an effective range $\sim 1/\mv \lapp 0.25 $ fm for $\mv \rapp \mro$.
Consequently, their matrix elements will carry a non-negligible sensitivity
to the short-distance part of the nuclear wave function. In principle, one
could account for this sensitivity by using Monte Carlo methods and
variational ground state wave functions to evaluate the two-body matrix
elements [21]. For purposes of setting the scale of the MEC contribution,
however, it is sufficient to use the simpler harmonic oscillator wave function
and to account for short-range anti-correlations by including a
phenomenological correlation function, $g(r)$, in the integral over
relative co-ordinate, $r$. Following Ref.~[22], we take this function
to have the form
$$
g(r) = C\left[ 1- e^{-r^2/d^2}\right]\ \ \ , \eqno(9)
$$
where $C$ is a constant adjusted to maintain  the  wave function normalization
and the parameter $d=0.84$ fm is obtained by fitting the nuclear matter
correlation function  of Ref.~[23]. With this choice for $g(r)$, the
two-body matrix elements can be evaluated analytically, thereby making
transparent the physical parameters which govern the scale of these
matrix elements.

        Looking first at the limit of un-correlated wave functions
($g(r)\equiv 1$), we obtain for the sum of one- and two-body
contributions to the strange-quark Coulomb matrix element
$$
\rbra{\hbox{g.s.}}\Mhat_0(s)\rket{\hbox{g.s.}}\Bigr\vert_{q\to 0} = \tau\left[
\lambda_1\rhostr + \lambda_{2\rm a}\mustr+\lambda_{2\rm b}\gropi\right]
\ \ \ ,\eqno(10)
$$
where
$$
\eqalignno{\lambda_1&=2/\sqrt{\pi}&(11{\rm a})\cr
         &\cr
         \lambda_{2\rm a}&\approx-\sum_{V=\rho,\omega}
         (1+\kappa_\sst{V})\left[{\sqrt{2}g_\sst{V}^2
         \over 8\pi^2}\right]\left[1-{5\over (\mv b)^2}+\cdots\right]\left(
         {\mv\over \mn}\right){{\cal N}_V\over(\mv b)^3}&(11{\rm b})\cr
         &\cr
         \lambda_{2\rm b}&={2\sqrt{2}\over 9\pi^2}\grnn\gpnn(1+\kappa_\rho)
         {{\cal N}_2\over \mro b}\left[{1\over (\mro b)^2-
         (\mpi b)^2}\right]&(11{\rm c})\cr
         &&\cr
        &\quad\times\left[(\mpi b)^2 I(\mpi b) - (\mro b)^2 I(\mro b)\right]
         \ \ \ ,&\cr}
$$
where
$$
I(m b) = 1-\sqrt{\pi\over 2}\ (m b) \ \exp\left({(m b)^2\over 2}\right)
        \ {\rm efrc}\left({ mb\over\sqrt{2}}\right) \ \ \ ,\eqno(12)
$$
and where $b$ is the oscillator parameter, ${\cal N}_{V,2 }$
are spin-isospin matrix
elements, and $g_\sst{V}$ is the vector meson-nucleon coupling.
{}From a fit to the $^4$He charge form factor, one obtains a value
for the oscillator parameter of $b=1.2$ fm [9].
In this case, one has $\mro b >> 1$ and
$$
(\mro b)^2 I(\mro b) \approx 1 - {3\over (\mro b)^2}+\cdots\ \ \ .\eqno(13)
$$
Numerically, the one-body matrix element gives $\lambda_1\approx 1.13$,
while use of the un-correlated wave function gives
$\lambda_{2\rm a}^{(\rho)} \approx -0.06$ and
$\lambda_{2\rm a}^{(\omega)} \approx -0.03$ for the $\rho$- and $\omega$-meson
pair currents, respectively, and $\lambda_{2\rm b}\approx -0.9$ for the
transition current. In obtaining these results,
we have used values for the couplings taken from Ref.~[21]: $\grnn = 2.6$,
$g_{\omega\sst{NN}} =
14.6$, $\kappa_\rho = 6.6$, and $\kappa_\omega = -0.12$
Although the scale of the un-correlated two-body
matrix elements is suppressed with respect to the one-body matrix element
by several powers of $1/(\mv b)$, this suppression is compensated in
the case of the transition current by the large values of the meson-nucleon
couplings and spin-isospin matrix element, ${\cal N}_2$. Thus, in a world
where anti-correlations became important only for $N-N$ separations
$<< 1/\mv$, the $^4$He strangeness radius could be as sensitive to
non-nucleonic strangeness as to the polarization of the nucleon's strange
sea ($|\lambda_{2\rm b}|\sim |\lambda_1|$).

	In the actual world, the $^4$He strangeness radius is dominated
by the one-body (single nucleon) contribution. Using the correlation
function of Eq.~(9) in computing the Coulomb matrix elements, we obtain
the same value of $\lambda_1$ as before and the following values for the
two-body contributions: $\lambda_{2\rm a}^{(\rho)} \approx -0.03$,
$\lambda_{2\rm a}^{(\omega)}\approx -0.015$, and $\lambda_{2\rm b}
\approx -0.06$. Whereas the pair-current contributions are reduced by
a factor two from their un-correlated values, the transition current
term is an order of magnitude smaller. The latter, more significant
suppression arises because the transition current matrix element involves
the difference of two operators,  whose ranges are set, respectively,
by $1/\mpi$ and $1/\mro$ (see Eq.~(7)). Short-range correlations reduce the
value of the second operator's matrix element but do not seriously affect the
first, so that the degree of cancellation between the two is enhanced. This
cancellation is somewhat sensitive to the values of $b$ and $d$ employed,
with $\lambda_{2\rm b}$ ranging from $\approx -0.1\to -0.01$ as $d$
is increased from $\sim 0.8 \to 1.0$ fm for $b=1.2$ fm. In contrast, the
values of $\lambda_{2\rm a}^{(\rho,\ \omega)}$ are stable with respect to
this variation.

        In terms of the PV $^4$He asymmetry, it is the ratio of the matrix
element in Eq.~(10) to the isoscalar EM Coulomb matrix element which governs
the second term in Eq.~(2). Since the numerator of this term is already
of ${\cal O}(\tau)$, we need retain only the $\tau=0$ part of the denominator
when extracting the nuclear strangeness radius from a low-$|Q^2|$ measurement
of the asymmetry. Consequently, we set the denominator $=\GETEZ(0)\lambda_1
=\lambda_1/2$. The resulting ratio is
$$
{\rbra{\hbox{g.s.}}\Mhat_0(s)\rket{\hbox{g.s.}}\over
\rbra{\hbox{g.s.}}\Mhat_0(T=0)\rket{\hbox{g.s.}}}\biggr\vert_{q\to 0}
=\quad 2\tau\rhostr\left[1+ \left({\lambda_{2\rm a}\over\lambda_1}
\right) \mustr + \left({\lambda_{2\rm b}\over \lambda_1}\right)\gropi
\right]+ {\cal O}(\tau^2)\ \ \ .\eqno(14)
$$
{}From these results we conclude that the one-boson-exchange
MEC's should not seriously impact the extraction of $\rhostr$ from the helium
asymmetry, unless $\mustr$ and $\gropi$ are an order of magnitude  larger
in scale than $\rhostr$. The latter possibility seems unlikely, since
a variety of calculations give $|\mustr|\lapp|\rhostr|$ [11-16]. Similarly,
an estimate of $\gropi$ using a $\phi$-dominance model and the measured
width for $\phi\to\rho\pi$ gives $|\gropi|\approx 0.26$ [24], which is of
the same magnitude or smaller than most estimates
for the magnitude of $\rhostr$.
\vskip 24pt
\centerline{\bf ACKNOWLEDGEMENTS}
\medskip
It is a pleasure to thank Jose Goity and Rocco Schiavilla for useful
discussions.
\bigskip
\centerline{\bf REFERENCES}
\medskip
\item{1.}L. A. Ahrens, \etal, \PRD{35} (1987) 785.
\medskip
\item{2.}D. B. Kaplan and A. Manohar, \NPB{310} (1988) 527.
\medskip
\item{3.}R. D. McKeown, \PLB{219} (1989) 140; MIT-Bates proposal \# 89-06,
R. D. McKeown and D. H. Beck, spokespersons.
\medskip
\item{4.}D. H. Beck, \PRD{39} (1989) 3248.
\medskip
\item{5.}S. J. Pollock, \PRD{42} (1990) 3010.
\medskip
\item{6.}J. Napolitano, \PRC{43} (1991) 1473.
\medskip
\item{7.}M. J. Musolf and T. W. Donnelly, \NPA{546} (1992) 509.
\medskip
\item{8.}M. J. Musolf and T. W. Donnelly, \ZPC{57} (1993) 559.
\medskip
\item{9.}M. J. Musolf, T. W. Donnelly, J. Dubach, S. J. Pollock,
S. Kowalski, and E. J. Beise, CEBAF Theory preprint \# TH-93-11 (1993).
\medskip
\item{10.}G. T. Garvey, W. C. Louis, and D. H. White, \PRC{48} (1993) 761.
\medskip
\item{11.}R. L. Jaffe, \PLB{229} (1989) 275.
\medskip
\item{12.}B. R. Holstein in {\it Parity Violation in Electron Scattering},
Proceedings of the Caltech Workshop, E. J. Beise and R. D. McKeown, Eds.,
World Scientific, 1990, p. 27.
\medskip
\item{13.}N. W. Park, J. Schechter, and H. Weigel, \PRD{43} (1991) 869.
\medskip
\item{14.}W. Koepf, E. M. Henley, and S. J. Pollock, \PLB{288} (1992) 11.
\medskip
\item{15.}M. J. Musolf and M. Burkardt, CEBAF Theory preprint \# TH-93-1
(1993).
\medskip
\item{16.}T. D. Cohen, H. Forkel, and M. Nielsen, Univ. of Maryland preprint
\# 93-217 (1993).
\medskip
\item{17.}CEBAF proposal \# 91-004, E. J. Beise, spokesperson.
\medskip
\item{18.}CEBAF proposal \# 91-010, J. M. Finn and P. A. Souder, spokespersons.
\medskip
\item{19.}CEBAF proposal \# 91-017, D. H. Beck, spokesperson.
\medskip
\item{20.}M. J. Musolf and T. W. Donnelly, to be published.
\medskip
\item{21.}R. Schiavilla, V. R. Pandharipande, and D.O. Riska, \PRC{41} (1990)
309.
\medskip
\item{22.}J. Dubach, J. H. Koch, and T. W. Donnelly, \NPA{271} (1976) 279.
\medskip
\item{23.}E. J. Moniz and G. D. Nixon, \AoP{67} (1971) 58.
\medskip
\item{24.}J. Goity and M. J. Musolf, to be published.
\bigskip
\centerline{\bf FIGURE CAPTIONS}
\medskip
\noindent{\bf Fig. 1.} \lq\lq Pair-current" (a,b) and \lq\lq meson-in-flight"
(c)
meson-exchange current (MEC) processes contributing to nuclear matrix elements
of the electromagnetic (EM) and vector strange-quark currents. Here, $N$ and
$N'$
denote two nucleons, $M$ a meson, and the crossed circle either the EM or
vector
strange-quark current operators. The process of Fig. 1c does not contribute
to the nuclear strangeness radius or isoscalar EM charge form factor.
\medskip
\noindent{\bf Fig. 2.} Isobar (a) and transition current (b)
MEC processes contributing the the nuclear
strangeness radius. The notation is identical to that of Fig. 1,
with $N^\ast$
denoting a nucleon resonance and $(M, M')$ any two mesons.
\vfill
\eject
\end

%
%
%

\def\gboxit#1{\hbox{\vrule\vbox{\hrule\kern3pt\vtop
{\hbox{\kern3pt#1\kern3pt}
\kern3pt\hrule}}\vrule}}

\def\ttilde#1{\raise2ex\hbox{${\scriptscriptstyle(}\!
\sim\scriptscriptstyle{)}$}\mkern-16.5mu #1}
\def\dddots#1{\raise1ex\hbox{$^{\ldots}$}\mkern-16.5mu #1}
\def\pp#1#2{\raise1.5ex\hbox{${#2}$}\mkern-17mu #1}
\def\upleftarrow#1{\raise2ex\hbox{$\leftarrow$}\mkern-16.5mu #1}
\def\uprightarrow#1{\raise2ex\hbox{$\rightarrow$}\mkern-16.5mu #1}
\def\upleftrightarrow#1{\raise1.5ex\hbox{$\leftrightarrow$}\mkern-16.5mu #1}
\def\bx#1#2{\vcenter{\hrule \hbox{\vrule height #2in \kern #1in\vrule}\hrule}}

\def\squiggle#1{\lower1.5ex\hbox{$\sim$}\mkern-14mu #1}

\def\narrower{\advance\leftskip by\parindent \advance\rightskip by\parindent}


\def\mbox#1#2{\vcenter{\hrule width#1in\hbox{\vrule height#2in
   \hskip#1in\vrule height#2in}\hrule width#1in}}
\def\eqsquare #1:#2:{\vcenter{\hrule width#1\hbox{\vrule height#2
   \hskip#1\vrule height#2}\hrule width#1}}
\def\inbox#1#2#3{\vcenter to #2in{\vfil\hbox to #1in{$$\hfil#3\hfil$$}\vfil}}
\def\strutdepth{\dp\strutbox}
\def\marbul{\strut\vadjust{\kern-\strutdepth\specialbul}}
\def\specialbul{\vtop to \strutdepth{
    \baselineskip\strutdepth\vss\llap{$\bullet$\qquad}\null}}
\def\Bcomma{\lower6pt\hbox{$,$}}    
\def\bcomma{\lower3pt\hbox{$,$}}    

\def\sl{\scrsf}

\def\updots{\mathinner{\mskip 1mu\raise 1pt\hbox{.}
    \mskip 2mu\raise 4pt\hbox{.}\mskip 2mu
    \raise 7pt\vbox{\kern 7pt\hbox{.}}\mskip 1mu}}

\def\square{\kern1pt\vbox{\hrule height 1.2pt\hbox{\vrule width 1.2pt\hskip 3pt
   \vbox{\vskip 6pt}\hskip 3pt\vrule width 0.6pt}\hrule height 0.6pt}\kern1pt}
\def\ssquare{\kern1pt\vbox{\hrule height .6pt\hbox{\vrule width .6pt\hskip 3pt
   \vbox{\vskip 6pt}\hskip 3pt\vrule width 0.6pt}\hrule height 0.6pt}\kern1pt}
\def\lege{\hbox{$ {     \lower.40ex\hbox{$>$}
                   \atop \raise.20ex\hbox{$<$}
                   }     $}  }

\def\rege{\hbox{$ {     \lower.40ex\hbox{$<$}
                   \atop \raise.20ex\hbox{$>$}
                   }     $}  }

\def\lapp{\hbox{$ {     \lower.40ex\hbox{$<$}
                   \atop \raise.20ex\hbox{$\sim$}
                   }     $}  }
\def\rapp{\hbox{$ {     \lower.40ex\hbox{$>$}
                   \atop \raise.20ex\hbox{$\sim$}
                   }     $}  }

\def\tridots{\hbox{$ {     \lower.40ex\hbox{$.$}
                   \atop \raise.20ex\hbox{$.\,.$}
                   }     $}  }
\def\Times{\times\hskip-2.3pt{\raise.25ex\hbox{{$\scriptscriptstyle|$}}}}

\def\rightonleft{\hbox{$ {     \lower.40ex\hbox{$\longrightarrow$}
                   \atop \raise.20ex\hbox{$\longleftarrow$}
                   }     $}  }

\def\pmb#1{\setbox0=\hbox{$#1$}%
\kern-.025em\copy0\kern-\wd0
\kern.05em\copy0\kern-\wd0
\kern-.025em\raise.0433em\box0 }

%
%
\font\fivebf=cmbx5
\font\sixbf=cmbx6
\font\sevenbf=cmbx7
\font\eightbf=cmbx8
\font\ninebf=cmbx9
\font\tenbf=cmbx10

\font\bfmone=cmbx10 scaled\magstep1

\font\sevenit=cmti7
\font\eightit=cmti8
\font\nineit=cmti9
\font\tenit=cmti10

\font\itmone=cmti10 scaled\magstep1

\font\fiverm=cmr5
\font\sixrm=cmr6
\font\sevenrm=cmr7
\font\eightrm=cmr8
\font\ninerm=cmr9
\font\tenrm=cmr10

\font\rmmone=cmr10 scaled\magstep1

\def\fontone{\def\rm{\fcm0\rmmone}%
  \textfont0=\rmmone \scriptfont0=\tenrm \scriptscriptfont0=\sevenrm
  \textfont1=\itmone \scriptfont1=\tenit \scriptscriptfont1=\sevenit
  \def\it{\fcm\itfcm\itmone}%
  \textfont\itfcm=\itmone
  \def\bf{\fcm\bffcm\bfmone}%
  \textfont\bffcm=\bfmone \scriptfont\bffcm=\tenbf
   \scriptscriptfont\bffcm=\sevenbf
  \tt \ttglue=.5em plus.25em minus.15em
  \normalbaselineskip=25pt
  \let\sc=\tenrm
  \let\big=\tenbig
  \setbox\strutbox=\hbox{\vrule height10.2pt depth4.2pt width\z@}%
  \normalbaselines\rm}



\font\ninerm=cmr9
\font\eightrm=cmr8
\font\sixrm=cmr6

\font\ninei=cmmi9
\font\eighti=cmmi8
\font\sixi=cmmi6
\skewchar\ninei='177 \skewchar\eighti='177 \skewchar\sixi='177

\font\ninesy=cmsy9
\font\eightsy=cmsy8
\font\sixsy=cmsy6
\skewchar\ninesy='60 \skewchar\eightsy='60 \skewchar\sixsy='60

\font\ninebf=cmbx9
\font\eightbf=cmbx8
\font\sixbf=cmbx6

\font\ninett=cmtt9
\font\eighttt=cmtt8

\hyphenchar\tentt=-1 
\hyphenchar\ninett=-1
\hyphenchar\eighttt=-1

\font\ninesl=cmsl9
\font\eightsl=cmsl8

\font\nineit=cmti9
\font\eightit=cmti8


\newskip\ttglue
\def\tenpoint{\def\rm{\fcm0\tenrm}%
  \textfont0=\tenrm \scriptfont0=\sevenrm \scriptscriptfont0=\fiverm
  \textfont1=\teni \scriptfont1=\seveni \scriptscriptfont1=\fivei
  \textfont2=\tensy \scriptfont2=\sevensy \scriptscriptfont2=\fivesy
  \textfont3=\tenex \scriptfont3=\tenex \scriptscriptfont3=\tenex
  \def\it{\fcm\itfcm\tenit}%
  \textfont\itfcm=\tenit
  \def\sl{\fcm\slfcm\tensl}%
  \textfont\slfcm=\tensl
  \def\bf{\fcm\bffcm\tenbf}%
  \textfont\bffcm=\tenbf \scriptfont\bffcm=\sevenbf
   \scriptscriptfont\bffcm=\fivebf
  \def\tt{\fcm\ttfcm\tentt}%
  \textfont\ttfcm=\tentt
  \tt \ttglue=.5em plus.25em minus.15em
  \normalbaselineskip=16pt
  \let\sc=\eightrm
  \let\big=\tenbig
  \setbox\strutbox=\hbox{\vrule height8.5pt depth3.5pt width\z@}%
  \normalbaselines\rm}

\def\ninepoint{\def\rm{\fcm0\ninerm}%
  \textfont0=\ninerm \scriptfont0=\sixrm \scriptscriptfont0=\fiverm
  \textfont1=\ninei \scriptfont1=\sixi \scriptscriptfont1=\fivei
  \textfont2=\ninesy \scriptfont2=\sixsy \scriptscriptfont2=\fivesy
  \textfont3=\tenex \scriptfont3=\tenex \scriptscriptfont3=\tenex
  \def\it{\fcm\itfcm\nineit}%
  \textfont\itfcm=\nineit
  \def\sl{\fcm\slfcm\ninesl}%
  \textfont\slfcm=\ninesl
  \def\bf{\fcm\bffcm\ninebf}%
  \textfont\bffcm=\ninebf \scriptfont\bffcm=\sixbf
   \scriptscriptfont\bffcm=\fivebf
  \def\tt{\fcm\ttfcm\ninett}%
  \textfont\ttfcm=\ninett
  \tt \ttglue=.5em plus.25em minus.15em
  \normalbaselineskip=11pt
  \let\sc=\sevenrm
  \let\big=\ninebig
  \setbox\strutbox=\hbox{\vrule height8pt depth3pt width\z@}%
  \normalbaselines\rm}

\def\eightpoint{\def\rm{\fcm0\eightrm}%
  \textfont0=\eightrm \scriptfont0=\sixrm \scriptscriptfont0=\fiverm
  \textfont1=\eighti \scriptfont1=\sixi \scriptscriptfont1=\fivei
  \textfont2=\eightsy \scriptfont2=\sixsy \scriptscriptfont2=\fivesy
  \textfont3=\tenex \scriptfont3=\tenex \scriptscriptfont3=\tenex
  \def\it{\fcm\itfcm\eightit}%
  \textfont\itfcm=\eightit
  \def\sl{\fcm\slfcm\eightsl}%
  \textfont\slfcm=\eightsl
  \def\bf{\fcm\bffcm\eightbf}%
  \textfont\bffcm=\eightbf \scriptfont\bffcm=\sixbf
   \scriptscriptfont\bffcm=\fivebf
  \def\tt{\fcm\ttfcm\eighttt}%
  \textfont\ttfcm=\eighttt
  \tt \ttglue=.5em plus.25em minus.15em
  \normalbaselineskip=9pt
  \let\sc=\sixrm
  \let\big=\eightbig
  \setbox\strutbox=\hbox{\vrule height7pt depth2pt width\z@}%
  \normalbaselines\rm}

\font\scrsf=cmssi10                

\font\ninebf=cmbx9




\def\gboxit#1{\hbox{\vrule\vbox{\hrule\kern3pt\vtop
{\hbox{\kern3pt#1\kern3pt}
\kern3pt\hrule}}\vrule}}

\def\ttilde#1{\raise2ex\hbox{${\scriptscriptstyle(}\!
\sim\scriptscriptstyle{)}$}\mkern-16.5mu #1}
\def\dddots#1{\raise1ex\hbox{$^{\ldots}$}\mkern-16.5mu #1}
\def\siton#1#2{\raise1.5ex\hbox{$\scriptscriptstyle{#2}$}\mkern-17.5mu #1}
\def\pp#1#2{\raise1.5ex\hbox{${#2}$}\mkern-17mu #1}
\def\upleftarrow#1{\raise2ex\hbox{$\leftarrow$}\mkern-16.5mu #1}
\def\uprightarrow#1{\raise2ex\hbox{$\rightarrow$}\mkern-16.5mu #1}
\def\upleftrightarrow#1{\raise1.5ex\hbox{$\leftrightarrow$}\mkern-16.5mu #1}
\def\bx#1#2{\vcenter{\hrule \hbox{\vrule height #2in \kern #1in\vrule}\hrule}}

\def\squiggle#1{\lower1.5ex\hbox{$\sim$}\mkern-14mu #1}

\def\narrower{\advance\leftskip by\parindent \advance\rightskip by\parindent}

\def\onsim{\hbox{$ {     \lower.40ex\hbox{$\sim$}
                   \atop \raise.20ex\hbox{$+$}
                   }     $}  }

\def\simon{\hbox{$ {     \lower.40ex\hbox{$+$}
                   \atop \raise.20ex\hbox{$\sim$}
                   }     $}  }


\def\mbox#1#2{\vcenter{\hrule width#1in\hbox{\vrule height#2in
   \hskip#1in\vrule height#2in}\hrule width#1in}}
\def\eqsquare #1:#2:{\vcenter{\hrule width#1\hbox{\vrule height#2
   \hskip#1\vrule height#2}\hrule width#1}}
\def\inbox#1#2#3{\vcenter to #2in{\vfil\hbox to #1in{$$\hfil#3\hfil$$}\vfil}}
\def\strutdepth{\dp\strutbox}
\def\marbul{\strut\vadjust{\kern-\strutdepth\specialbul}}
\def\specialbul{\vtop to \strutdepth{
    \baselineskip\strutdepth\vss\llap{$\bullet$\qquad}\null}}
\def\Bcomma{\lower6pt\hbox{$,$}}    
\def\bcomma{\lower3pt\hbox{$,$}}    

\def\sl{\scrsf}

\def\updots{\mathinner{\mskip 1mu\raise 1pt\hbox{.}
    \mskip 2mu\raise 4pt\hbox{.}\mskip 2mu
    \raise 7pt\vbox{\kern 7pt\hbox{.}}\mskip 1mu}}

\def\square{\kern1pt\vbox{\hrule height 1.2pt\hbox{\vrule width 1.2pt\hskip 3pt
   \vbox{\vskip 6pt}\hskip 3pt\vrule width 0.6pt}\hrule height 0.6pt}\kern1pt}
\def\ssquare{\kern1pt\vbox{\hrule height .6pt\hbox{\vrule width .6pt\hskip 3pt
   \vbox{\vskip 6pt}\hskip 3pt\vrule width 0.6pt}\hrule height 0.6pt}\kern1pt}
\def\lege{\hbox{$ {     \lower.40ex\hbox{$>$}
                   \atop \raise.20ex\hbox{$<$}
                   }     $}  }

\def\rege{\hbox{$ {     \lower.40ex\hbox{$<$}
                   \atop \raise.20ex\hbox{$>$}
                   }     $}  }

\def\tridots{\hbox{$ {     \lower.40ex\hbox{$.$}
                   \atop \raise.20ex\hbox{$.\,.$}
                   }     $}  }
\def\Times{\times\hskip-2.3pt{\raise.25ex\hbox{{$\scriptscriptstyle|$}}}}

\def\rightonleft{\hbox{$ {     \lower.40ex\hbox{$\longrightarrow$}
                   \atop \raise.20ex\hbox{$\longleftarrow$}
                   }     $}  }

\def\pmb#1{\setbox0=\hbox{$#1$}%
\kern-.025em\copy0\kern-\wd0
\kern.05em\copy0\kern-\wd0
\kern-.025em\raise.0433em\box0 }